# Using Clustering to Understand Intra-city Warming in Heatwaves: Insights into Paris, Montreal, and Zurich

**Yongling Zhao[1]\*, Dominik Strebel[1], Dominique Derome[2], Igor Esau[3], Qi Li[4], Jan Carmeliet[1]**

[1] Department of Mechanical and Process Engineering, ETH Zürich, Zürich, Switzerland

[2] Department of Civil and Building Engineering, Université de Sherbrooke, Québec, Canada

[3] Department of Physics and Technology, The Arctic University of Norway

[4] School of Civil and Environmental Engineering, Cornell University, Ithaca, USA

   Corresponding author: Yongling Zhao (yozhao@ethz.ch)

**Abstract**

We introduce a novel methodological advancement by clustering paired near-surface air temperature with the planetary boundary layer height (PBLH) to characterize intra-city clusters for analytics. To illustrate this approach, we analyze three heatwaves (HW): the 2019 HW in Paris, the 2018 HW in Montreal, and the 2017 HW in Zurich. We assess cluster-based characteristics before, during, and after heatwave events. Using the objective hysteresis model, we determine the overall strength coefficient of the hysteresis loop between ground storage flux and all-wave downward radiative flux, ranging from 0.414 to 0.457 for urban clusters and from 0.126 to 0.157 for rural clusters during the heatwave periods. Across all cities, we observe a consistent refueling-restoration mode in the cumulative ground heat flux as the heatwaves progress. Future developments of this proposed two-component clustering approach, with the integration of more influential physics, will offer a more comprehensive characterization of cities for urban climate analytics.

## 1 Introduction

With over half of the global population now residing in urban areas, the urban heat island (UHI) phenomenon – wherein urban areas are warmer than their surrounding rural areas (Oke, 1981) – requires a profound understanding of its underlying physics (Estrada et al., 2017). It also necessitates the development of both mitigation and adaptation strategies (Li et al., 2014; Krayenhoff et al., 2018; Debnath et al., 2023; Zhao et al., 2023). Differences in either surface temperature (Kalnay et al., 2003; Manoli et al., 2020; Li et al., 2023) or near-surface air temperature (Kubilay et al., 2020; Chiu et al., 2022) between urban and rural areas are commonly determined to quantify the intensity of the surface or canopy UHI (Oke, 1976; Oke, 2006). It has been widely observed that the temporal variation in UHI intensity tends to be more pronounced at night than during the day (Oke, 1981; Li et al., 2020). Insights into UHI intensity from these studies often correlate it with factors such as wind speed, anthropogenic heat emission, heat absorption by urban materials, and a variety of biophysical processes (e.g. by vegetation), especially those associated with vegetation within the urban canopy layer (Arnfield, 2003; Song et al., 2015; Wang et al., 2018; Tan et al., 2023).

The UHI effect emerges from the intricate interaction between surface heat fluxes and atmospheric characteristics (Kropfli et al., 1978; Oke, 1981; Hildebrand et al., 1984; Miao et al., 2009). Above the ground, the planetary boundary layer (PBL) is the region where various forms of heat are mixed, transported, and distributed. The excessive heat from cities, a contributing factor to UHI, is retained within the atmospheric layer. Kropfli et al. (1978) observed that roll vortices, once formed, seemed to be locked in place in reaction to the urban environment. Hildebrand and Ackerman[16] reported enhanced vertical fluxes of temperature, moisture, and along-wind stress in urban areas. They inferred that urban roughness's impacts are less important compared to influence of urban heating, especially under moderate wind conditions. A convective velocity scale correlating the surface vertical heat flux and atmospheric turbulence was proposed as $w_* = (\frac{g}{T}\overline{T'w'}Z_i)^{1/3}$, where $\overline{T'w'}$ is the surface vertical heat flux, and $Z_i$ is a representative length scale taken as the height of the inversion of the PBL above the ground (Deardorff et al., 1982). The link between urban heat and PBL was further observed in field measurements in smaller urban settings like Christchurch (Tapper, 1990). Using data from Toulouse, Hidalgo et al. (2008) revealed that urban breezes can transport heat from the city center's lower atmospheric to the higher atmospheric layers downwind. A recent LES-WRF study by Zhu et al. (2016) highlighted the influence of both momentum roughness length and thermal roughness length on the surface heat flux and the PBL over urban landscapes.

During heatwave events (HWs), the PBL is influenced by both synoptic-scale atmospheric circulation and land-atmosphere interactions (Miralles et al., 2014; Zhang et al., 2020). During the day, the planetary boundary layer height (PBLH) is notably greater during HWs compared to non-HWs, though the difference becomes less pronounced at night. The influence of high-pressure zones, combined with soil moisture-temperature feedback, has been recognized in mega-heatwaves (Miralles et al., 2014). Furthermore, complex synergistic interactions between HWs and the UHIs phenomenon have been observed (Li et al., 2013; Zhao et al., 2018).

When the PBLH is shallow, temperature variations can be strongly amplified (Esau et al., 2012), given the PBL's small heat capacity. The PBL inherently connects surface heat fluxes with atmospheric characteristics (Davy et al., 2016; Lu et al., 2023). Lin et al. (2008) noted that anthropogenic heat, manifesting as increased surface heat flux, can elevate the PBLH by several hundred meters. Pal et al. (2012) found that UHI intensity has a more pronounced effect on PBLH at night than during the day. Varentsov et al. (2023) recently observed the dependence of PBLH to UHI in a sub-Arctic city. In the framework of the surface energy budget and atmospheric characteristics, Chiu et al. (2022) proposed an analytical solution for predicting both urban heat and moisture islands.

Studies on the PBL and heat islands often begin by defining the locations of urban and rural areas, typically using land types or a city's development plan as reference points. While this method facilitates comparative studies between urban and rural characteristics, the distinction between the two does not rely on inherent physical characteristics. Rather, it introduces subjectivity, which can sometimes obscure the definitions of urban and rural areas. This approach can be inadequate in at least three frequent occasions: (1) urban and rural classifications may not be updated promptly in rapidly developing regions, making the available land type data less accurate; (2) the definitions of urban and rural areas can vary across different regions and countries; and (3) high levels of anthropogenic emissions in what are classified as rural areas can challenge conventional definitions of rural and urban spaces. Such inconsistencies complicate the development of universal models and generalizable understanding.

We advocate for a data-driven, objective approach that relies on the intrinsic characteristics of UHI and atmospheric factors to categorize intra-city clusters for urban climate research. We first construct a paired dataset, combining two-meter air temperature and PBLH for each computational grid. This dataset then undergoes machine learning-assisted clustering to discern clusters with distinct urban or rural features. As a validation of this methodology, we applied it to three different cities: Paris, Montreal, and Zurich, which have populations of approximately 2.2 million, 1.8 million, and 0.4 million, respectively. Urban heat islands during three recent heatwaves in 2019, 2018, and 2017 served as the climatic conditions against which urban and rural cluster characterizations were validated. Using these identified clusters, we conducted analyses on diurnal PBL, surface heat budget, and hysteresis.

**2 Setup of WRF simulations and two-component Clustering**

2.1 Heatwave data

We conducted mesoscale simulations of heatwaves for representative cities in periods before, during, and after the heatwaves to create spatial and temporal datasets for clustering analysis. The near-surface air temperature from the simulation of each city was validated with available weather station data. The cities examined are Paris, Montreal, and Zurich. We simulated the heatwaves that occurred in 2019, 2018, and 2017 for the three cities.

2.2 Configuration of WRF simulations

Mesoscale simulations were performed using the Advanced Research WRF (WRF-ARW, version 4.2) (Skamarock et al., 2021). The global ERA-5 dataset from the European Center for Medium-Range Weather Forecasts (ECMWF) was used for the initial and boundary conditions for all the simulated cases (Hersbach et al., 2020). The dataset was obtained from the Copernicus Climate Data service at 3-hour intervals in the model as well as on surface levels on a 0.25° × 0.25° grid. The MODIS-IGBP land cover dataset and GMTED2010 30-arc-second layer as the digital elevation model were adopted in the simulations (Friedl et al., 2002; Danielson et al., 2011).

The resolution of the mesoscale domains was set to 1000 m × 1000 m for all the simulations. The extent of the domains varies according to the spatial size of the simulated cities. The latitudinal extent differed between 229 and 251 km, and longitudinally between 251 and 391 km. The number of vertical levels was set to 101 and manually configured to provide a regular spacing close to the boundary layer. No nesting was performed to save computational power, and the area of interest was placed at the center of the simulation domain. All simulations were configured with sufficient spin-up time before the start of the respective heatwave. The Mellor-Yamada-Janjic (MYJ) scheme was adopted for PBL parameterization. The unified NOAH LSM and Eta Similarity scheme (Janjic, 1994) were used in the surface layer model as urban parametrization.

2.3 Two-component clustering

Two-component clustering (i.e., temperature and PBLH) was performed in three steps: $S_1$, $S_2$, and $S_3$, as schematically illustrated in Figure 1(a). First, in $S_1$, we paired the air temperature T2 at 2 m with the critical $PBLH$ on a grid level at hourly resolution, using six nocturnal time instants (1 am to 6 am) over three consecutive days during a heatwave event. This was given the conditions of low wind speeds and a distinct urban-rural contrast (e.g. Pal et al., 2012). The dataset is represented as $\boldsymbol{D} = D_{TH(i,j,t)}$, where $i, j$ denote the grid index, and $t$ denotes the time instant. Here, $PBLH$ is estimated by scanning the local Richardson number from the ground to the height, which results in a critical Richardson number $Ri_{cr}$ of 0.25 (Davy et al., 2016). The corresponding $PBLH$ was calculated using $PBLH = g^{-1} Ri_{cr}(U^2 + V^2)T_{vp0}/(T_{vp} - T_{vp0})$, where $g$ is the surface gravity (9.807 $ms^{-2}$); $U$ and $V$ represent the local wind speed; $T_{vp0}$ is the virtual potential temperature at the surface, and $T_{vp}$ is the virtual potential temperature at the observation height. The air temperature at 2 m ($T2$) was directly adopted from the urban parameterization simulation results.

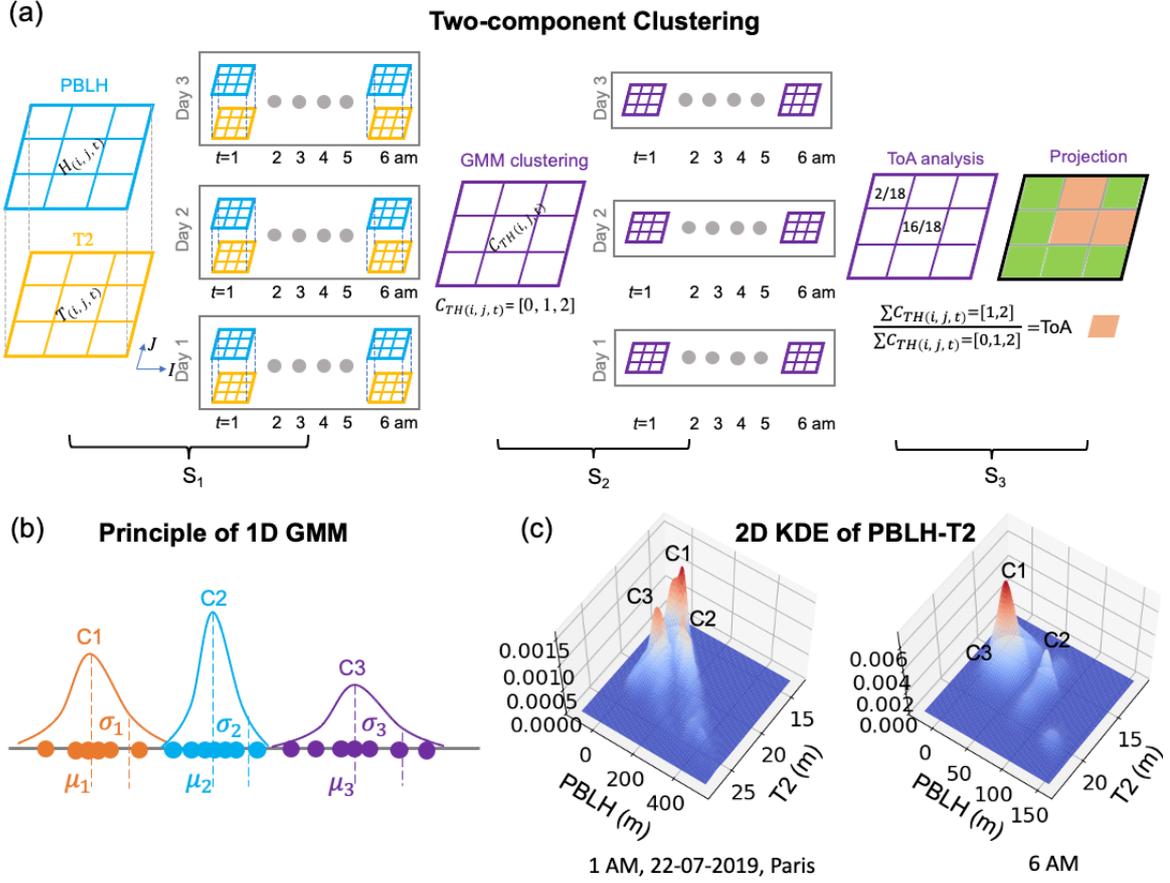

**Figure 1.** The GMM-based clustering to determine urban and rural clusters. (a) Framework for the two-component GMM clustering, (b) principle of 1D GMM, and (c) 2D Kernel density estimation of the two-component (PBLH and T2) dataset at two time instants.

Second, in $S_2$, we performed clustering analysis using Gaussian mixture models (GMM) because of its flexibility in forming clusters with unequal variances. The principle behind the application of GMM to a one-dimensional (1D) dataset is schematically illustrated in figure 1(b). This assumes all data points are generated from a mixture of a finite number of Gaussian distributions with an unknown set of means ($\mu$) and covariances ($\sigma$). Similarly, to applying GMM to a 1D univariate Gaussian distribution, clustering for the two-dimensional (2D) dataset ***D*** can be performed assuming the multivariate Gaussian distribution of ***D***, following a probability density function:

$$p(d; \mu, \sigma) = \frac{1}{2\pi^{n/2} \lceil \sigma \rceil^{1/2}} \exp\left(-\frac{1}{2}(d-\mu)^T \sigma^{-1} (d-\mu)\right) \quad (1)$$

where $p(\cdot)$ denotes density functions; $\boldsymbol{D} \sim \mathbb{N}(\mu, \sigma)$ has a mean matrix $\mu \in \boldsymbol{M}^n$ and a covariance matrix $\sigma \in \boldsymbol{CV}^n$ (Wade, 2023). Examples showing the two-dimensional kernel distribution estimation (KDE) of the dataset ***D*** observed at two instants are presented in figure 1(c). This clearly suggests the presence of three density functions, as indicated by C1, C2 and C3.

An expectation-maximization (EM) algorithm is used to perform maximum likelihood estimation with a convergence threshold of $10^{-3}$. Each component is estimated with its own general covariance matrix, and the prior on weights is specified with a Dirichlet process (DP) prior. The number of clusters is chosen to be three to represent expected rural, suburban and urban features. Depending on the nature of the problem of interest, other number of clusters can be used.

Third, in $S_3$, a threshold-of-appearance (ToA) was proposed and applied at the grid scale to determine if it is part of the urban cluster (i.e., C1). For a given grid, the times of appearance of the clustering result being part of the urban cluster (C1), relative to all the clustering possibilities (C1, C2 and, C3), is counted as ToA:

$$\text{ToA} = \frac{\sum D_{TH(i,j,t)}|C1}{\sum D_{TH(i,j,t)}|C1, C2, C3} \quad (2)$$

where C1 denotes the dataset $D_{TH(i,j,t)}$ being recognized as a grid $(i, j)$ of the cluster C1 at the time instant $t$. In our clustering analysis, the ToA in the range of 0.8~0.9 was observed to effectively identify urban areas that are in good agreement with the corresponding land type. A smaller ToA may result in an overestimated cover of urban areas, while a larger ToA usually leads to an underestimation. The use of the ToA aims to mitigate the relatively weak coupling between PBLH and T2 due to short-period strong horizontal convection during the day. Similar observations were made in the field measurements by Pal et al. (2012), where the urban-rural contrast in PBLH was observed to occur about 85% of the time of the cycle. Zhu et al. (2016) also noted that urban circulation diminishes when the wind speed exceeds 5 m/s. In subsequent analyses, suburban and urban clusters are agglomerated as a simplification within the scope of this study.

## 3 Results

3.1 Deep heat dome above the urban cluster

In Kropfli et al. (1978), it was discover that a distinct horizontal roll mode of convection was 'locked' to the thermal features of an urban area. This type of convection, arising from turbulent vertical heat flux, leads to a heat dome above cities (e.g. Fan et al., 2017; Zhang et al., 2023). Here, we initiate our analysis by examining the coupling of the PBLH and thermal features. Figure 2 presents the instantaneous spatial profile of PBLH, near-surface air temperature, and the corresponding urban cluster for Paris, as determined by the two-component clustering approach. As shown in Figure 2a, the PBLH, extending several hundred meters deep, is distinctly positioned above the Paris center. The spatial profile of the near-surface air temperature is less homogenous, with temperatures ranging from 38 °C to 22°C, still distinctly shows an unequivocal Paris center (Figure 2b). By pairing the air temperature with PBLH, the domain above the terrain can be classified into two clusters: one represents an urban cluster (in brown) and the other signifies a rural cluster (in deep blue), as shown in Figure 2c. These two clusters largely reflect the morphology of Paris compared to the land-use types, implying the 'locked' head islands effect. Similar heat domes above the urban areas of Zurich and Montreal are also evident in our simulation results, which can be inferred in Figure 3 presented in next section.

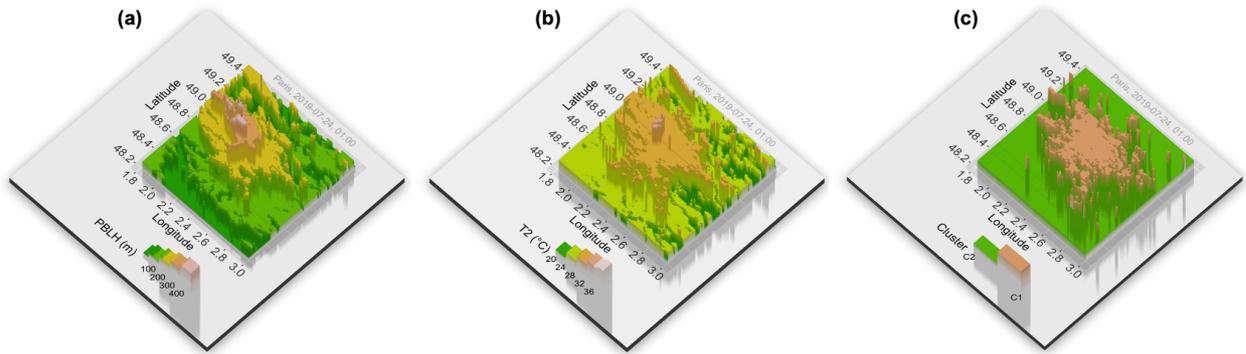

**Figure 2**. Instantaneous Coupling of PBLH and T2. (a) Spatial profile of PBLH over Paris at a local time 00:00 on July 24, 2019, (b) the corresponding 2-meter air temperature, and (c) urban and rural clusters distinguished by the coupling of the nighttime PBLH and T2 over three consecutive days (22-24 July, 2019).

The presence of a persistent heat dome can be significantly disturbed or altered by high wind speeds. Both Hildebrand et al. (1984) and Zhu et al. (2016) noted that when the geostrophic wind speed reaches approximately 5–6 m s$^{-1}$, the urban-centered circulation diminishes. An urban heat dome typically manifests when wind speeds are around 2 m s$^{-1}$ (Hidalgo et al., 2008). During heatwaves, decreased wind speeds in urban areas, as highlighted by Li et al. (2016), tend to foster the formation of heat domes.

3.2 Cluster-based diurnal near-surface temperature and PBLH

The 'locked' urban clusters, along with their temperature and PBLH, are shown in Figure 3. As illustrated in Figure 3($a_1$-$c_1$), two-component clustering of the PBLH and 2-m air temperature effectively distinguishes urban (in brown) and rural clusters (in green) across the three cities. This representation highlights the unique urban morphologies of each city: a centralized urban area in Paris, urban zones encircled by rivers in Montreal, and patches of urban regions adjacent to the lake in Zurich. In the clustering process, the threshold of appearance (ToA) was set at 16 out of 18, as this yielded satisfactory clustering results. A similar observation was made by Pal et al. (2012), where the consistency in PBLH variation between urban and rural areas was approximately 85%. Notably, water bodies are generally distinguishable from urban clusters. However, some exceptions at the grid level likely arise from local large-scale convective flows that alter the coupling between air temperature and the planetary boundary layer.

The clustering approach utilized to differentiate urban and rural clusters in cities offers potential for spatial ensemble analytics. First, we analyzed the ensemble averages of 2-m air temperature (Figure 3$a_2$-$c_2$) and PBLH (Figure 3$a_3$-$c_3$) for urban and rural clusters during heatwave days. A striking observation is the correlation between elevated air temperatures in urban clusters (orange) and an increase in PBLH.

Notably, while the air temperature difference between urban and rural clusters is more pronounced at night, PBLH differences are more distinct during the day. The nighttime air temperature difference is up to 4–5 °C between urban and rural clusters in both Paris and Montreal, whereas daytime PBLH discrepancies extend to several hundred meters. During a 4-day lidar measurement of PBLH in March 2011 under clear sky conditions in Paris, the differences in PBLH between urban and rural areas spanned from approximately -200 to 600 m during the day and 0 to 200 m at night (Pal et al., 2012). The data highlighted a robust associate between PBLH and near-surface air temperature, deemed to be undeniable for both urban and suburban areas. This correlation proved stronger for urban dataset compared to rural ones.

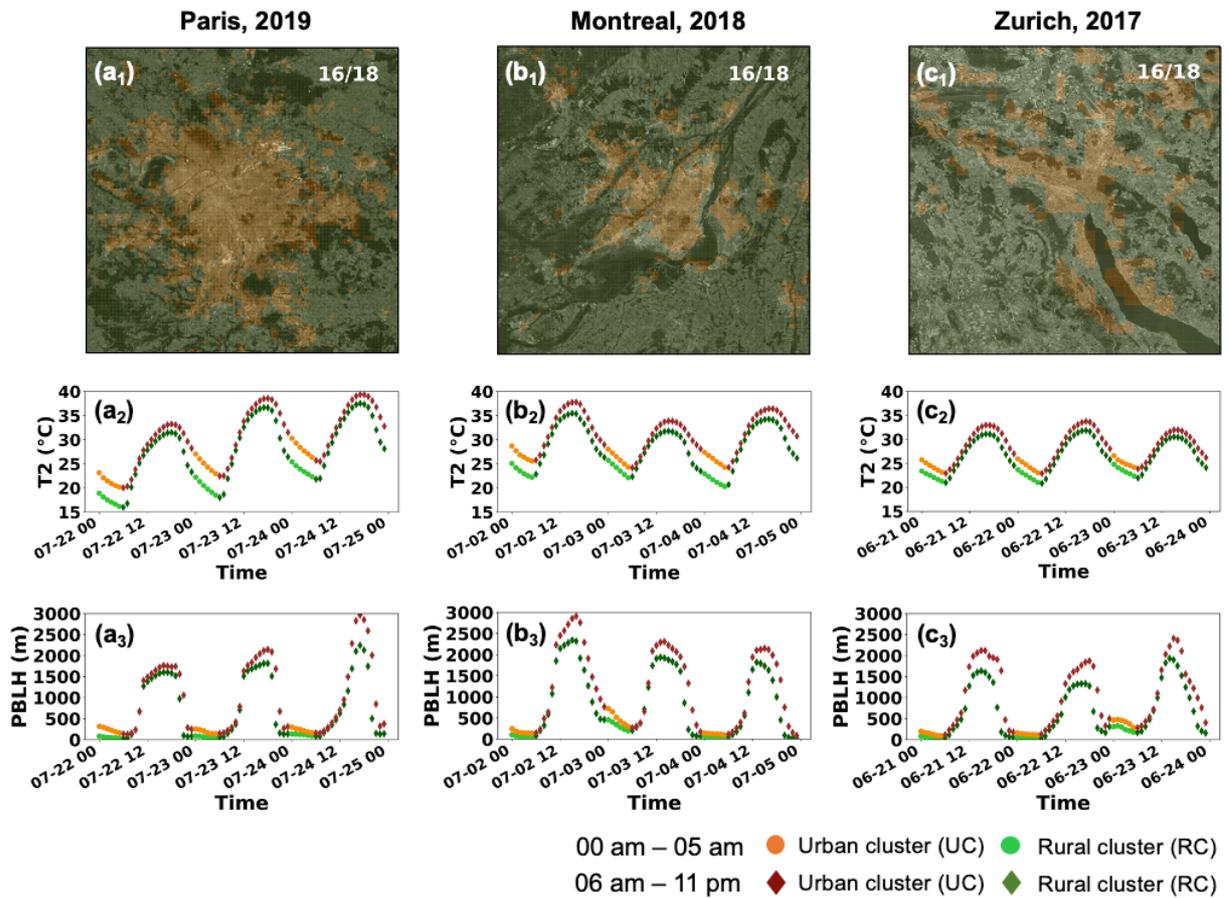

**Figure 3.** Urban and rural clusters and cluster-based analytics. ($a_1$ – $c_1$) Urban and rural clusters denoted by brown and green in Paris, Montreal, and Zurich, ($a_2$ – $c_2$) cluster-based 2-m air temperature in three consecutive days in respective heatwaves occurred in 2019, 2018, and 2017, and ($a_3$ – $c_3$) cluster-based PBLH. Data points for nighttime are highlighted in deep colors.

3.3 Hysteresis of the near-surface temperature and PBLH

The observed diurnal variations of urban-rural contrasts in air temperature and PBLH, as well as the associated phase lag, suggest a hysteresis relationship between near-surface air temperature and PBLH. Here, we examine the hysteresis of the paired PBLH and 2-m air temperature for both rural and urban clusters during three key periods: before, during, and after heatwaves. The direction of the hysteresis loop (DoHL) is indicated. Figure 4 clearly shows that the hysteresis trajectories for the three cities vary remarkably for both rural and urban clusters during heatwaves. One of the most distinct findings is the pronounced upward expansion of the hysteresis trajectories for urban clusters on the diagram of air temperature versus PBLH. This indicates a concurrent deepening of the PBLH and intensification of the air temperature during heatwaves, as depicted in Fig.4 ($a_2$, $b_2$, $c_2$). Conversely, it is noteworthy that the hysteresis trajectories seen pre-heatwaves tend to revert during post-heatwaves for both cluster types, as evidenced in Fig.4 ($a_3$, $b_3$, $c_3$). Manoli et al. (2020) highlighted the existence of a hysteresis effect in surface urban heat islands on a seasonal scale, which also exhibits variability across different climatic zones. The dynamics of the hysteresis observed in the three characteristic periods remains elusive, suggesting that the underlying physics might be more intricate than solely the interplay between air temperature and PBLH.

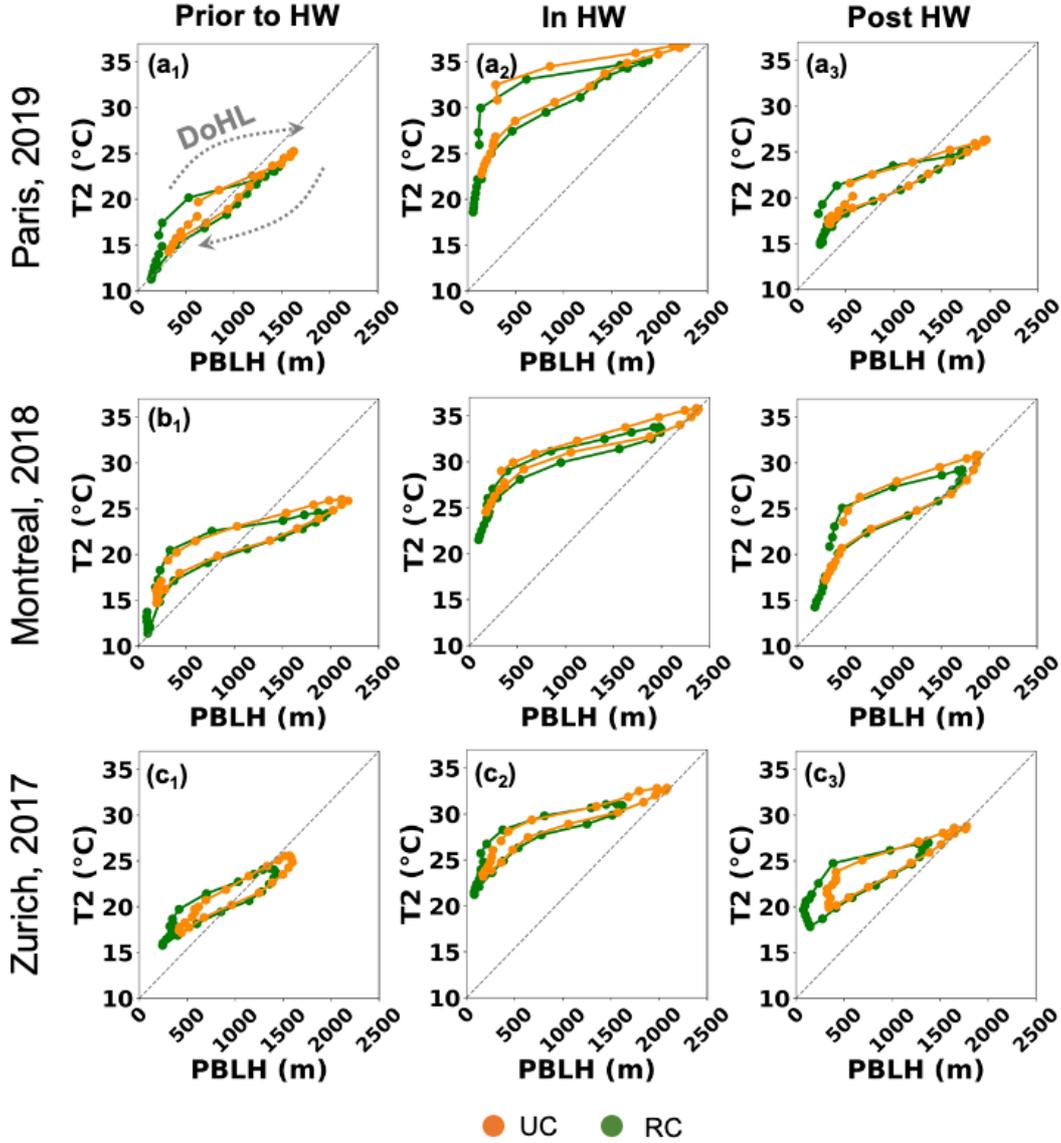

**Figure 4.** Hysteresis of the coupled planetary boundary layer height and 2-m air temperature prior to, during and post heatwaves for rural and urban areas, respectively. a.1 – a.3 16-18 June, 21-23 July, and 25-27 July of Zurich in 2017 for a period of three consecutive days prior to, in and post heatwaves, respectively. b.1 – b.3 14-16 July, 2019 22-24 July, and 27-29 July of Paris in 2019. c.1 – c.3 25-27 June, 2-4 July and 7-9 July of Montreal in 2018.

3.4 Accumulative characteristics of surface heat fluxes

The surface heat fluxes are typically studied based on predefined urban and rural areas (Grimmond et al., 1999; Hidalgo et al., 2008). Here we analyze the surface heat fluxes based on rural and urban clusters, and the cluster-based surface energy balance can be written as Yu et al. (2021):

$$R + Q_{An} = H + L + G + Q_{Ad} \qquad (3)$$

where $R$ denotes the net, downward all-wave radiative flux, and $Q_{An}$ stands for the anthropogenic heat flux. $H$, $L$ and $G$ denote the upward surface sensible heat flux, the upward surface latent heat flux due to evaporation, and the storage heat flux of ground. $Q_{Ad}$ denotes the heat due to advection between rural and urban clusters, which is assumed to be negligible (Chiu et al., 2022). $Q_{An}$ is not considered in our WRF simulations that use the NOAH LSM.

Figure 5a$_1$-c$_1$ shows the distribution of four heat flux components, resolved hourly over a 3-day period, in three cities (Paris, Montreal, and Zurich) during three distinct phrases: before (denoted by subscript 'Pri'), during (subscript 'In'), and after (subscript 'Pos') a heatwave. The components include the ground heat flux ($G$), latent heat flux at the surface ($L$), upward sensible heat flux ($H$), and incoming radiation ($R$).

In all three periods, it is evident that the most significant disparity in heat flux between rural and urban clusters lies in the distribution of the ground heat flux $G$ and latent heat flux $L$. The urban cluster consistently displays a broader distribution of the former and a narrower band of the latter, implying highly inhomogeneous ground heat flux and more uniform latent heat. This difference arises because the ground heat flux is influenced by a wider range of factors, such as surface albedo, emissivity, and urban canopy wind, compared to the factors affecting latent heat, leading to a more varied distribution of $G$. Similarly, a broader distribution of sensible heat flux $H$ is observed in urban clusters across the three cities compared to their rural counterparts. As anticipated, the net radiation on the ground surface showed minimal disparity between the rural and urban clusters. Notably, the three cities exhibit similar distributions for the various surface fluxes, suggesting analogous heat transport mechanisms are at work before, during, and after the heatwaves.

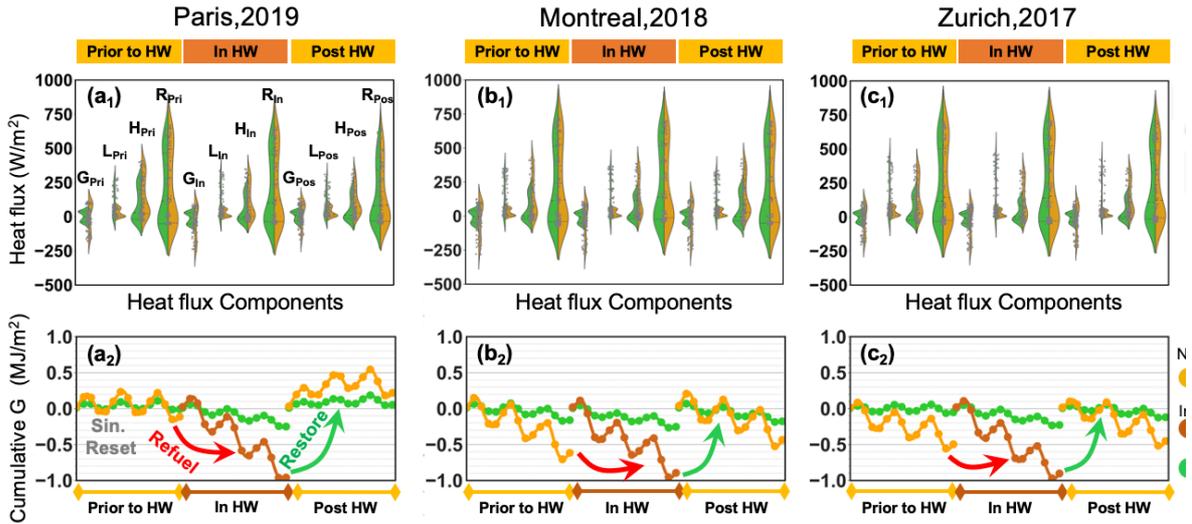

**Figure 5.** Rural and urban cluster-based surface energy budget and hysteresis analysis. (a$_1$-c$_1$): Distribution of surface heat flux components in a 3-day time window prior to, during, and post a heatwave, and (a$_2$-c$_2$) hourly integration of the ground heat flux ($G$) in the 3-day time windows.

However, the distribution of these 3-day heat fluxes cannot fully account for the amplified heat island intensity. A timewise integration of the ground heat flux G over their respective 3-day time windows is performed to explore possible mechanisms. This is illustrated in Figure 5a$_2$-c$_2$, where negative quantities imply downward fluxes. Before the occurrence of the heatwave in Paris, the ground heat fluxes (see Figure 5a$_2$) for both urban and rural clusters exhibit a clear diurnal sinusoidal pattern, denoted by 'Sin. Reset', allowing them to be 'reset' to 'net zero.' However, during the heatwave, this 'reset' cannot be maintained, leading to rapid intensification and an increase in the downward ground heat flux. After the heatwave, the diurnal features of the ground heat flux were restored. This alternating pattern in ground heat flux is significant and can be termed the refueling-restoration mode. This mode is also evident during the heatwaves observed in Montreal and Zurich (Figure 5b$_2$ and c$_2$). Notably, in both Montreal and Zurich, the intensification of the ground heat flux already starts to develop in urban clusters before the heatwaves, while the diurnal sinusoidal reset is still distinct in their rural clusters.

To further examine the hysteresis of the storage heat flux in relation to the all-wave downward radiative flux, the objective hysteresis model (OHM) proposed by Grimmond et al. (1991) is adopted, which is presented for clarity:

$$G = a_1 R + a_2 \frac{\partial R}{\partial t} + a_3 \qquad (4)$$

where $a_1$ indicates the overall strength of the dependence of the storage heat flux on all-wave radiative heat flux, $a_2$ describes the phase relations between the two fluxes, and $a_3$ indicates the relative timing when $G$ and $R$ turn negative. The derivative of the all-wave downward radiative heat flux is approximated as:

$$\frac{\partial R}{\partial t} = (R_{t+1} - R_{t-1})/2 \qquad (5)$$

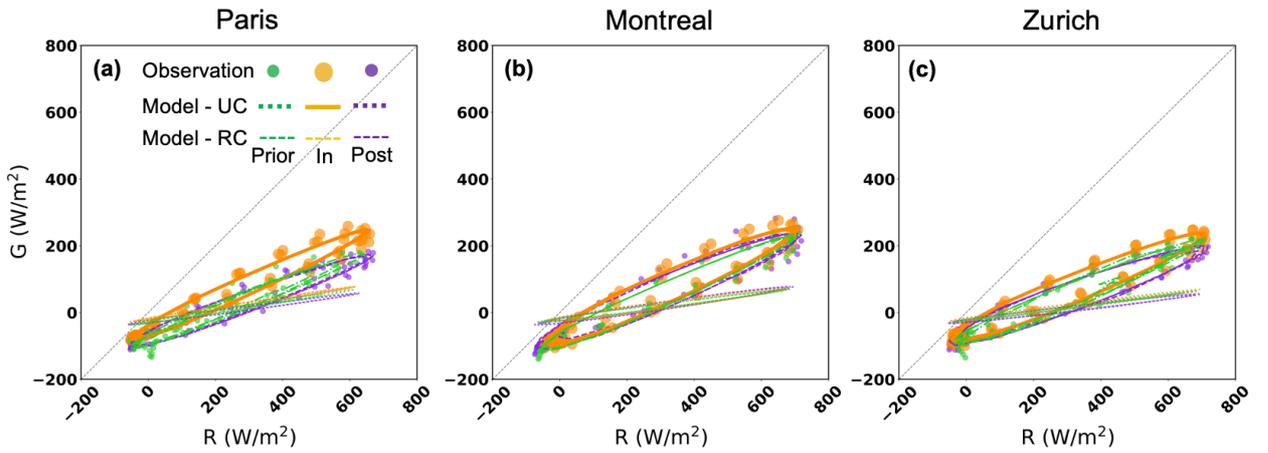

**Figure 6.** Rural and urban cluster-based hysteresis analysis. (a-c) The hysteresis loops of the storage heat flux G and the downward heat flux R in three time windows of the urban and rural clusters.

The hysteresis loops for both urban and rural clusters during the three characteristic time windows are shown in Figure 6a-c. The most pronounced observation is the substantially stronger hysteresis exhibited by the urban clusters compared to the rural ones across all three cities. This observation is corroborated by the OHM coefficient $a_1$ listed in Table A1. Additional coefficients of the objective hysteresis model for the heatwaves in Paris, Montreal, and Zurich can also be found in Table A1. Notably, during the heatwave days, a moderate strengthening of the hysteresis is observed for the urban clusters in these cities.

## 5 Conclusions

This study first introduces a data-driven clustering approach to characterize rural and urban clusters, leveraging the coupling of nighttime PBLH and 2-m air temperature. The clustering results from Paris, Montreal, and Zurich confirm that rural and urban regions can be distinguished well using two-component clustering. Furthermore, this cluster-based approach allows for deeper, physics-oriented investigations. The diurnal variations in PBLH and 2-m air temperature within these urban and rural clusters that correspond to conventionally defined urban and rural areas can be well diagnosed.

The distinct coupling of the nighttime 2-m air temperature and PBLH – which encapsulates the connection between the canopy layer's vertical heat flux and the characteristics of the PBL – can potentially be used as a two-dimensional criterion. This criterion is valuable for predicting the onset and decay of urban heat islands effects during heatwaves, as illustrated using a hysteresis diagram. Finally, an analysis of the surface energy budget based on urban and rural clusters revealed a notable intensification-restoration mode in ground storage heat flux.

While the two-component clustering method highlighted in this work effectively distinguishes intra-city heat-related phenomena, it is still primitive. Important factors such as wind speed and extensive vegetation cover are not accounted for. Specifically, when wind speeds exceed 5 m s$^{-1}$, the current two-component approach may not distinguish between urban and rural clusters. In these situations, a multi-component clustering could better capture the intricate interactions between ground heat and the PBL. Additionally, when using more than two clusters, it becomes possible to focus on the climate dynamics at the suburban or neighborhood level. However, the impact of choosing a specific number of clusters on the results warrants careful consideration.

**Acknowledgement**: Y Zhao would like to acknowledge the discussion with Dr. Davy Richard at Nansen Environmental and Remote Sensing Center and Bjerknes Centre for Climate Research.

**Appendix**

Table A1. OHM coefficients determined based on urban and rural clusters prior to, during and post heatwaves.

| City | Cluster | Period | OHM coefficients | | | $R^2$ |
| | | | $a_1$ | $a_2$ (h) | $a_3$ (W m-2) | |
|---|---|---|---|---|---|---|
| Paris | UC | Prior | 0.370 | 0.254 | -76.594 | 0.984 |
| | | In | **0.457** | 0.245 | -52.812 | 0.985 |
| | | Post | 0.401 | 0.210 | -80.293 | 0.955 |
| | RC | Prior | 0.135 | 0.050 | -26.887 | 0.944 |
| | | In | 0.157 | 0.032 | -19.145 | 0.951 |
| | | Post | 0.158 | 0.046 | -28.456 | 0.927 |
| Montreal | UC | Prior | 0.420 | 0.359 | -68.280 | 0.974 |
| | | In | **0.450** | 0.370 | -68.985 | 0.985 |
| | | Post | 0.443 | 0.263 | -83.159 | 0.985 |
| | RC | Prior | 0.147 | 0.082 | -24.570 | 0.922 |
| | | In | 0.133 | 0.077 | -19.224 | 0.953 |
| | | Post | 0.132 | 0.046 | -20.275 | 0.957 |
| Zurich | UC | Prior | 0.361 | 0.333 | -65.161 | 0.982 |
| | | In | **0.414** | 0.335 | -59.772 | 0.988 |
| | | Post | 0.413 | 0.269 | -75.682 | 0.974 |
| | RC | Prior | 0.116 | 0.055 | -24.843 | 0.947 |
| | | In | 0.126 | 0.044 | -19.154 | 0.943 |
| | | Post | 0.127 | 0.030 | -22.935 | 0.938 |